\documentclass[reprint,twocolumn,superscriptaddress,showpacs,showkeys,preprintnum bers,amsmath,amssymb,floatfix,aps,prl,longbibliography]{revtex4-1}

\usepackage{graphicx}
\usepackage{bm}

\usepackage{placeins}
\usepackage{graphics}
\usepackage{color}
\usepackage{hyperref}
\usepackage{multirow}
\usepackage{blindtext}

\usepackage{pdfpages}

\makeatletter
\AtBeginDocument{\let\LS@rot\@undefined}
\makeatother  

\begin{document}


\title{Bilayer graphene encapsulated within monolayers of WS$_2$ or Cr$_2$Ge$_2$Te$_6$: \\Tunable proximity spin-orbit or exchange coupling
}

\author{Klaus Zollner}
	\email{klaus.zollner@physik.uni-regensburg.de}
	\affiliation{Institute for Theoretical Physics, University of Regensburg, 93053 Regensburg, Germany}	
\author{Jaroslav Fabian}
	\affiliation{Institute for Theoretical Physics, University of Regensburg, 93053 Regensburg, Germany}
\date{\today}

\begin{abstract}
Van der Waals (vdW) heterostructures consisting of bilayer graphene (BLG) encapsulated within monolayers of strong spin-orbit semiconductor WS$_2$ or ferromagnetic semiconductor Cr$_2$Ge$_2$Te$_6$ (CGT), are investigated. 
By performing realistic first-principles calculations we capture the essential BLG band structure features, including layer- and sublattice-resolved proximity spin-orbit or exchange couplings. For different relative twist angles (0 or 60$^{\circ}$) of 
the WS$_2$ layers, and the magnetizations (parallel or antiparallel) of the CGT layers, with respect to BLG, the low energy bands are found and characterized by a series of fit parameters of model Hamiltonians. These effective models are then employed to investigate the tunability of the relevant energy dispersions by a gate
field. For WS$_2$/BLG/WS$_2$ encapsulation we find that twisting allows to turn off the spin splittings away from the $K$ points, due to
opposite proximity-induced valley-Zeeman couplings in the two sheets of BLG. Close to the $K$ points the electron spins are polarized out of the plane. This polarization can be flipped by applying a gate field. As for magnetic CGT/BLG/CGT structures, we realize the recently proposed spin-valve effect, whereby a gap
opens for antiparallel magnetizations of the CGT layers. Furthermore, we find that for the antiferromagnetic orientation the electron 
states away from $K$ have vanishingly weak proximity exchange, while the states close to $K$ remain spin polarized in the presence of an electric field. The induced magnetization can be flipped by changing the gate field. These findings should be useful for spin transport, spin filtering, and spin relaxation anisotropy studies of BLG-based vdW heterostructures. 
\end{abstract}

\maketitle

\section{Introduction}

Layered two-dimensional (2D) van der Waals (vdW)
materials have become indispensable for exploring new functionalities in electronics and spintronics \cite{Han2014:NN,Fabian2007:APS,Avsar2019:arxiv}. The fact that distinct 2D materials mutually influence each other in vdW heterostructures via proximity effects \cite{Zutic2019:MT} has opened venues for novel device designs at the nanoscale \cite{Gong2019:SC,Li2019:AM,Cortie2019:AFM}. 
It has been shown that several interactions, such as superconductivity \cite{Li2020:PRB,Moriya2020:PRB}, magnetism \cite{Zollner2016:PRB,Zollner2018:NJP,Zollner2019a:PRB,Zollner2020:PRB}, and spin-orbit coupling (SOC) \cite{Gmitra2015:PRB,Gmitra2016:PRB,Zollner2019b:PRB,Zollner2021:PSSB}, can be induced on demand, while
simultaneously each individual layer maintains its characteristic properties. Moreover, 
these proximity-induced spin interactions
can be further modulated by gating and twisting \cite{Zollner2019a:PRB,Song2018:NL,David2019:arxiv,Avsar2017:ACS,Ghiasi2019:NL,Benitez2020:NM,Luo2017:NL,Safeer2019:NL,Herlin2020:APL,Alsharari2018:PRB,Li2019:PRB}.

In this context, bilayer graphene (BLG) has emerged as a model playground for gate- and twist-tunable correlated physics \cite{Cao2018:Nat,Cao2018a:Nat,Arora2020:arxiv,Stepanov2020:Nat,Lu2019:Nat,Sharpe2019:SC,Ribeiro2018:SC, Saito2021:Nat,Serlin2020:S,Tschirhart2020:arxiv,Bultinck2020:PRL,Repellin2020:PRL,Zhang2019:PRR,Alavirad2020:PRB,Liu2021:PRB,Choi2019:NP,Lisi2021:NP,Balents2020:NP,Wolf2019:PRL} as well as for layer-dependent proximity-induced spin interactions \cite{Gmitra2017:PRL, Amann2021:arxiv, Zollner2020:PRL,Lin2021:arxiv,Wang2019:NL,Zollner2018:NJP, Island2019:Nat, Cardoso2018:PRL,Tiwari2021:PRL,Alsharari2018:PRB2}. 
Two sheets of graphene stacked at a small twist angle can become insulating, ferromagnetic \cite{Sharpe2019:SC,Bultinck2020:PRL,Alavirad2020:PRB}, or superconducting \cite{Cao2018:Nat,Cao2018a:Nat,Arora2020:arxiv,Stepanov2020:Nat,Lu2019:Nat} under certain filling factors of the Moir\'{e} Brillouin zone. 
By tuning the twist angle, one controls the interlayer coupling, thereby tailoring electronic and optical properties of BLG \cite{Nimbalkar2020:NML}.
In addition, also other microscopic details are highly important, such as strain, dielectric environment, and the alignment to encapsulation layers \cite{Sharpe2019:SC,Serlin2020:S,Stepanov2020:Nat,Saito2021:Nat,Balents2020:NP}.

Also proximity effects can induce spin interactions in BLG \cite{Zutic2019:MT,Song2018:NL,Omar2018:PRB,Omar2019:PRB,Gmitra2017:PRL,Gmitra2015:PRB,Gmitra2016:PRB,Zollner2018:NJP,Zollner2020:PRL,Zollner2019b:PRB,Zollner2016:PRB,Karpiak2019:arxiv,Khokhriakov2018:SA,Hoque2019:arxiv,Amann2021:arxiv,Herlin2020:APL,Avsar2017:ACS,Frank2016:PRB,Phong2017:2DM,Qiao2014:PRL,Leutenantsmeyer2016:2DM,Zihlmann2018:PRB,Ghiasi2017:NL,Jafarpisheh2018:PRB,Wang2015:PRL,Yang2013:PRL,Hallal2017:2DM,Wakamura2019:PRB,Chen2021:NP}. As theoretically predicted by first-principles calculations \cite{Gmitra2017:PRL} and recently confirmed via penetration field capacitance measurements \cite{Island2019:Nat} and mesoscopic transport \cite{Tiwari2021:PRL}, a transition-metal dichalcogenide (TMDC) in proximity to BLG strongly enhances the SOC of the adjacent graphene layer only. 
By applying a gate voltage, one can then fully electrically turn on and off the SOC of the BLG conduction electrons. Proximity effects even allow for swapping spin interactions (SOC and exchange) in vdW structures, such as 
BLG encapsulated between a ferromagnetic and a strong spin-orbit layer \cite{Zollner2020:PRL}. 
Such a "bottom-up" approach to vdW engineering, building more complex structures beyond bilayers, is just starting to reveal its enormous potential
for tailoring charge, spin, and optical properties of 2D materials. 

\begin{figure*}[!htb]
	\includegraphics[width=0.99\textwidth]{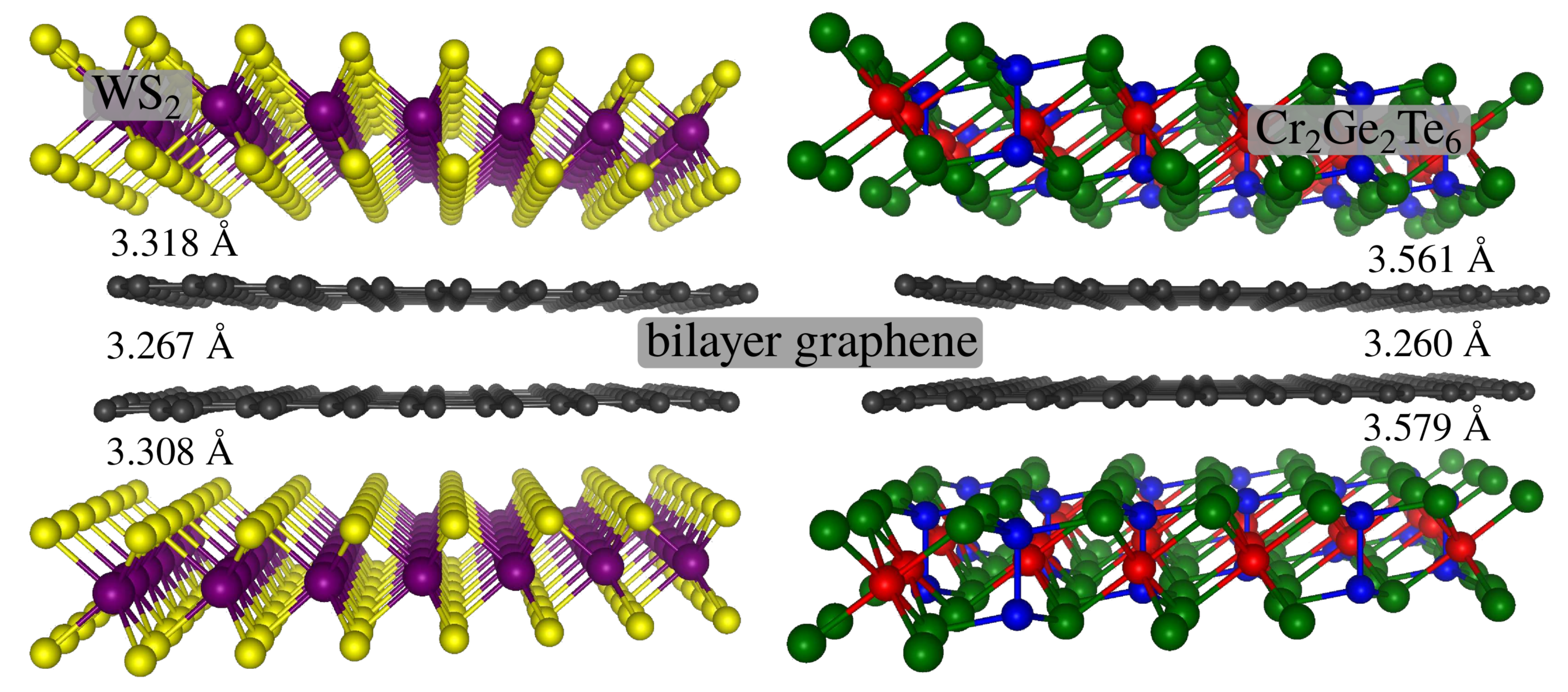}
	\caption{Geometries of the encapsulated BLG heterostructures. Left: WS$_2$ encapsulated BLG, Right: CGT encapsulated BLG. The relaxed interlayer distances are also indicated. 
 \label{Fig:Structure}}
\end{figure*}

Particularly interesting  are the recent experimental findings by Lin et al. \cite{Lin2021:arxiv}, who studied the combined effects of strong electronic correlations and SOC in WSe$_2$ proximitized magic-angle twisted BLG. Remarkably, the proximity Rashba and valley Zeeman SOC induces  orbital magnetism without the need for a rotational alignment to a hexagonal boron-nitride substrate~\cite{Sharpe2019:SC,Serlin2020:S}. In addition, WSe$_2$ can help to stabilize superconductivity in twisted BLG \cite{Arora2020:arxiv}, emphasizing the role of the dielectric environment. 
Also for the interpretation of such experimental findings it is useful 
to have a quantitative knowledge about proximity effects in BLG. For example, as demonstrated in Ref.~\cite{Island2019:Nat}, the measured penetration field capacitance of WSe$_2$ encapsulated BLG could be nicely related to model calculations, including proximity-induced valley-Zeeman SOC of opposite sign on the two graphene layers.

In this work, we consider BLG encapsulated either by monolayers of the strong spin-orbit semiconductor WS$_2$ or by monolayers of the ferromagnetic semiconductor Cr$_2$Ge$_2$Te$_6$ (CGT). 
In such vdW heterostructures, BLG preserves a great degree of autonomy of its electronic structure, but 
both graphene layers experience proximity effects (SOC by WS$_2$ or exchange coupling by CGT). 
In general, also other transition-metal dichalcogenides (MoS$_2$, MoSe$_2$, and WSe$_2$) or 2D magnets (CrI$_3$, MnPSe$_3$) are possible choices. Here, WS$_2$ and CGT serve as exemplary materials, inducing the spin interactions in the graphene layers.
In the case of WS$_2$ encapsulation, and when all layers are perfectly aligned ($0^{\circ}$ twist angle), we find that the induced SOC is of valley-Zeeman type ($\lambda_{\textrm{I}}^\textrm{A} \approx - \lambda_{\textrm{I}}^\textrm{B} \approx 1$~meV) and the same for both graphene layers. 
By twisting the top WS$_2$ layer by 
$60^{\circ}$, with respect to the underlying BLG/WS$_2$ structure, the valley-Zeeman SOC of the top graphene layer switches sign ($\lambda_{\textrm{I}}^\textrm{A} \rightarrow  - \lambda_{\textrm{I}}^\textrm{A}$, $\lambda_{\textrm{I}}^\textrm{B} \rightarrow  - \lambda_{\textrm{I}}^\textrm{B}$). 
Rotating the top WS$_2$ layer by $60^{\circ}$, we effectively switch the A and B sublattice of the underlying graphene layer.  
This has dramatic consequences for the low energy bands of the WS$_2$ encapsulated BLG, under zero external transverse electric field. The bands are spin-split, gapless, and $s_z$ spin-polarized for $0^{\circ}$ twist angle, while for $60^{\circ}$, they remain nearly unsplit and exhibit a band gap. 
The reason is the sublattice- and layer-polarized low energy band structure of BLG in combination with the short-range and layer-resolved proximity-induced valley-Zeeman SOC, that can be controlled by the twist angle. Our first-principles results explicitly prove the capacitance measurements on TMDC encapsulated BLG performed by Island et al. \cite{Island2019:Nat}.

Similarly, when BLG is encapsulated within two CGT monolayers, their individual magnetizations control the sign of the proximity-induced exchange coupling in the corresponding graphene layer. When both CGT layers have parallel magnetization, both graphene layers experience similar and uniform proximity exchange ($\lambda_{\textrm{ex}}^\textrm{A} \approx \lambda_{\textrm{ex}}^\textrm{B} \approx -3.5$~meV). 
By switching the magnetization direction of the top CGT layer, the proximity-induced exchange coupling of the adjacent graphene layer switches sign ($\lambda_{\textrm{ex}}^\textrm{A} \rightarrow  - \lambda_{\textrm{ex}}^\textrm{A}$, $\lambda_{\textrm{ex}}^\textrm{B} \rightarrow  - \lambda_{\textrm{ex}}^\textrm{B}$). 
Similar to before, the consequence for the BLG is that we can turn on and off its conductance by tuning the layer-resolved proximity exchange coupling at zero external transverse electric field. This realizes the recently proposed
spin-valve vdW heterostructure~\cite{Cardoso2018:PRL}.
A finite electric field can be used to tune the band gap, as well as spin and charge transport properties.

\section{Computational Details and Geometry}

In the following we consider BLG in Bernal stacking, which is encapsulated within two layers of WS$_2$ or CGT, see Fig. \ref{Fig:Structure}. 
Initial atomic structures are set up with the atomic simulation environment (ASE) \cite{ASE} and visualized with VESTA software \cite{VESTA}.
For the first-principles calculations we use $5 \times 5$ supercells of BLG,
$\sqrt{3} \times \sqrt{3}$ CGT supercells, and $4 \times 4$ supercells of WS$_2$.
In the case of WS$_2$ encapsulation, we stretch the lattice constant of BLG by roughly 2\% from $2.46~\textrm{\AA}$ to $2.5~\textrm{\AA}$ and 
the WS$_2$ lattice constant is compressed by about 1\% from $3.153~\textrm{\AA}$ \cite{Schutte1987:JSSC} to $3.125~\textrm{\AA}$.
In the case of CGT encapsulation, we keep the lattice constant of graphene unchanged at $a = 2.46~\textrm{\AA}$ and 
stretch the CGT lattice constant by roughly 4\% from $6.8275~\textrm{\AA}$ \cite{Carteaux1995:JPCM} to $7.1014~\textrm{\AA}$.

The electronic structure calculations and structural relaxations of 
the BLG-based heterostructures are performed by density functional theory 
(DFT) \cite{Hohenberg1964:PRB} with \textsc{Quantum ESPRESSO} \cite{Giannozzi2009:JPCM}.
Self-consistent calculations are performed with the $k$-point sampling of 
$12\times 12\times 1$ ($9\times 9\times 1$) in the case of CGT (WS$_2$) encapsulation to get converged results for the proximity-induced exchange (SOC). 
We use an energy cutoff for the charge density of $500$~Ry, and
the kinetic energy cutoff for wavefunctions is $60$~Ry for the scalar relativistic pseudopotentials 
with the projector augmented wave method \cite{Kresse1999:PRB} with the 
Perdew-Burke-Ernzerhof exchange correlation functional \cite{Perdew1996:PRL}.
In the case of CGT encapsulation, 
we perform open shell calculations that provide the 
spin polarized ground state and proximity exchange coupling. In addition,   
a Hubbard parameter of $U = 1$~eV is used for Cr $d$-orbitals, 
similar to recent calculations \cite{Gong2017:Nat, Zollner2018:NJP}.
In the case of WS$_2$ encapsulation, we use the relativistic versions of the pseudopotentials, to capture (proximity) SOC effects. 

For the relaxation of the heterostructures, we add
vdW corrections \cite{Grimme2006:JCC,Barone2009:JCC} and use 
quasi-newton algorithm based on trust radius procedure. 
Dipole corrections \cite{Bengtsson1999:PRB} are also included to get 
correct band offsets and internal electric fields.
In order to simulate quasi-2D systems, we add a vacuum of $20$~\AA, 
to avoid interactions between periodic images in our slab geometries.
Only the WS$_2$/BLG/WS$_2$ structure with $0^{\circ}$ twist angles, as well as the 
CGT/BLG/CGT structure with parallel magnetizations, are relaxed.
To determine the interlayer distances, the atoms of BLG and WS$_2$ 
are allowed to relax only in their $z$ positions 
(vertical to the layers), and the atoms of CGT are allowed to move in all directions,
until all components of all forces are reduced below $10^{-3}$~[Ry/$a_0$], where $a_0$ is the Bohr radius.
When we then twist the top WS$_2$ layer by $60^{\circ}$, or flip the magnetization of the top CGT layer, no further structural relaxation is performed.

The obtained interlayer distances are summarized in Fig.~\ref{Fig:Structure} and are similar to previous reports \cite{Gmitra2017:PRL, Zollner2018:NJP}.
Since we have assumed perfectly aligned individual layers, the full heterostructures still have $C_3$ symmetry after relaxation. In addition, the relaxed structures are nearly, but not fully symmetric regarding interlayer distances. The reason is that the precise atomic registries (stackings) of top and bottom encapsulation layer, with respect to their corresponding graphene sheet, are not exactly the same.

\section{Model Hamiltonian}

\begin{figure}[htb]
	\includegraphics[width=0.75\columnwidth]{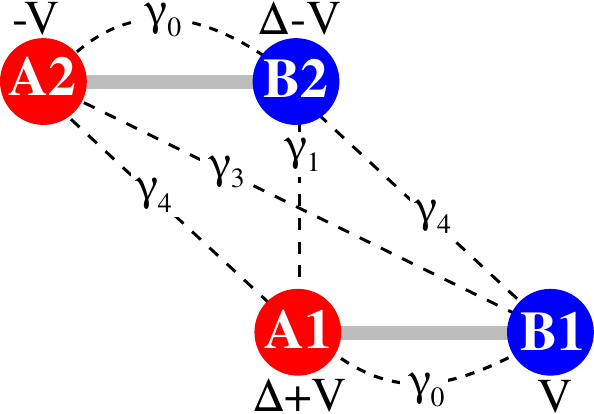}
	\caption{Schematic illustration of the BLG lattice, showing the relevant intra- and interlayer hoppings  $\gamma_j$, $j = \{ 0,1,3,4 \}$ (dashed lines). In addition, the lower (upper) graphene layer is placed in the potential $V$ ($-V$), while the asymmetry $\Delta$ arises due to $\gamma_1$.
 \label{Fig:BLG_scheme}}
\end{figure}

Here we present the Hamiltonian used to model the low energy bands of the encapsulated BLG structures. The basis states are
$|\textrm{C}_{\textrm{A1}}, \uparrow\rangle$, $|\textrm{C}_{\textrm{A1}}, \downarrow\rangle$, 
$|\textrm{C}_{\textrm{B1}}, \uparrow\rangle$, $|\textrm{C}_{\textrm{B1}}, \downarrow\rangle$, 
$|\textrm{C}_{\textrm{A2}}, \uparrow\rangle$, $|\textrm{C}_{\textrm{A2}}, \downarrow\rangle$, 
$|\textrm{C}_{\textrm{B2}}, \uparrow\rangle$, and $|\textrm{C}_{\textrm{B2}}, \downarrow\rangle$.
In this basis the Hamiltonian is (see also Refs. \onlinecite{Konschuh2012:PRB,Zollner2020:PRL})
\begin{widetext}
\begin{flalign}
\mathcal{H} = & \mathcal{H}_{\textrm{orb}} + \mathcal{H}_{\textrm{soc}}+\mathcal{H}_{\textrm{ex}}+\mathcal{H}_{\textrm{R}}+E_D,\\
\mathcal{H}_{\textrm{orb}} = & \begin{pmatrix}
\Delta+V & \gamma_0 f(\bm{k}) & \gamma_4 f^{*}(\bm{k}) & \gamma_1 \\
\gamma_0 f^{*}(\bm{k}) & V & \gamma_3 f(\bm{k}) & \gamma_4 f^{*}(\bm{k}) \\
 \gamma_4 f(\bm{k}) & \gamma_3 f^{*}(\bm{k}) & -V & \gamma_0 f(\bm{k}) \\
\gamma_1 & \gamma_4 f(\bm{k}) & \gamma_0 f^{*}(\bm{k}) & \Delta-V
\end{pmatrix} \otimes s_0,\\
 \mathcal{H}_{\textrm{soc}}+\mathcal{H}_{\textrm{ex}}+\mathcal{H}_{\textrm{R}} = &
\begin{pmatrix}
(\tau \lambda_{\textrm{I}}^\textrm{A1}-\lambda_{\textrm{ex}}^\textrm{A1})s_z & 2\textrm{i}\lambda_{\textrm{R1}}s_{-}^{\tau} & 0 & 0\\
-2\textrm{i}\lambda_{\textrm{R1}}s_{+}^{\tau} & (-\tau \lambda_{\textrm{I}}^\textrm{B1}-\lambda_{\textrm{ex}}^\textrm{B1})s_z & 0 & 0\\
0 & 0 & (\tau \lambda_{\textrm{I}}^\textrm{A2}-\lambda_{\textrm{ex}}^\textrm{A2})s_z & 2\textrm{i}\lambda_{\textrm{R2}}s_{-}^{\tau}\\
0 & 0 & -2\textrm{i}\lambda_{\textrm{R2}}s_{+}^{\tau} & (-\tau \lambda_{\textrm{I}}^\textrm{B2}-\lambda_{\textrm{ex}}^\textrm{B2})s_z
\end{pmatrix}.
\end{flalign}
\end{widetext}
Here, $\gamma_j$, $j = \{ 0,1,3,4 \}$, describe intra- and interlayer hoppings in BLG, as schematically illustrated in Fig.~\ref{Fig:BLG_scheme}. 
The parameter $\gamma_0$ is the nearest neighbor intralayer hopping, similar to monolayer graphene, while $\gamma_1$ is the direct interlayer hopping. The parameters $\gamma_3$ and $\gamma_4$ describe indirect hoppings between the layers. 
The vertical hopping $\gamma_1$ connects only 2 atoms and therefore it appears without structural function in the Hamiltonian. 
In contrast, the other hoppings couple an atom to three corresponding nearest neighbor partner atoms, hence they appear with structural function, where we use the linearized version, 
$f(\bm{k}) = -\frac{\sqrt{3}a}{2}(\tau k_x-\textrm{i}k_y)$, valid in the vicinity of the K points \cite{Kochan2017:PRB}. The graphene lattice constant is
$a$ and the Cartesian wave vector components $k_x$ and $k_y$ are measured with respect to $\pm K$ for the valley indices $\tau = \pm 1$. 
In addition, the lower (upper) graphene layer is placed in the potential $V$ ($-V$). The parameter $\Delta$ describes the asymmetry in the energy shift of the bonding and antibonding states, which arises due to the interlayer coupling $\gamma_1$.
The Pauli spin matrices are $s_i$, with $i = \{ 0,x,y,z \}$, and $s_{\pm}^{\tau} = \frac{1}{2}(s_x\pm \textrm{i}\tau s_y)$.

The parameters $\lambda_{\textrm{I}}$ ($\lambda_{\textrm{ex}}$) 
describe the proximity-induced intrinsic SOC (exchange) of the corresponding layer and sublattice atom 
($\textrm{C}_{\textrm{A1}}, \textrm{C}_{\textrm{B1}}, \textrm{C}_{\textrm{A2}}, \textrm{C}_{\textrm{B2}}$).
The intrinsic SOC parameters are also present in pristine BLG and on the order of 10~$\mu$eV \cite{Konschuh2012:PRB}. However, they will be strongly renormalized, because of the surrounding WS$_2$ layers providing an extra (on the meV scale) amount of SOC for the effective $p_z$-orbitals.
The exchange couplings are due to surrounding magnetic materials, giving rise to similar band splittings as the SOC parameters, but breaking time-reversal symmetry. 
The parameters $\lambda_{\textrm{R1}}$ and $\lambda_{\textrm{R2}}$ are the Rashba couplings of the two individual graphene layers.  
In pristine BLG, the two Rashba parameters have the same value but are opposite in sign, because the graphene layers feel each others presence.
When an electric field is applied across BLG, the Rashba couplings can become dissimilar. However, the surrounding WS$_2$ layers can strongly enhance the Rashba couplings, just as for monolayer graphene \cite{Gmitra2016:PRB}.

To capture doping effects from the calculations, we introduce another parameter $E_D$, which leads to an energy shift on the model band structure and we call it the Dirac point energy.
To extract the fit parameters form the DFT, we employ a least-squares routine, taking into account band energies, splittings, and spin expectation values. First, we extract the orbital parameters, $\gamma_j$, $\Delta$, and $V$. Once they are fixed, we extract the spin-orbit or exchange parameters, depending on the encapsulation layers. 

\section{Band Structure, Fit Results, and Gate Tunability}

\subsection{WS$_2$ encapsulated BLG}

\begin{figure*}[htb]
	\includegraphics[width=0.99\textwidth]{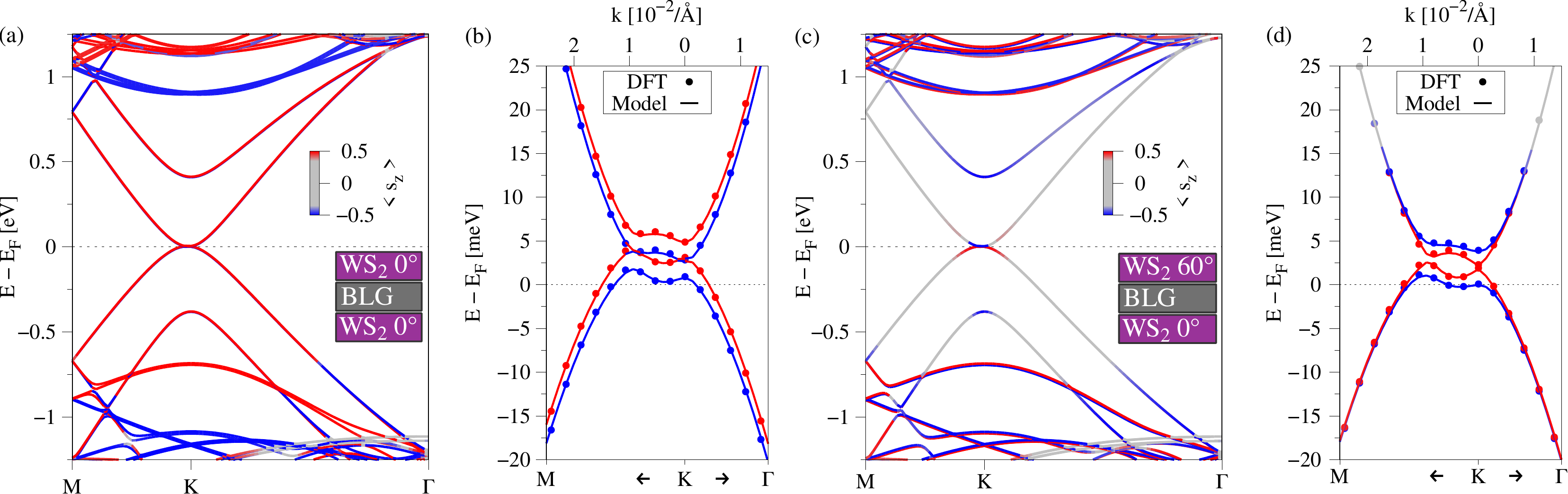}
	\caption{(a) Band structure of WS$_2$ encapsulated BLG along the $M-K-\Gamma$ path. The color of the bands corresponds to the $s_z$ spin expectation value. In the inset, we schematically illustrate the heterostructure, where both WS$_2$ layers have $0^{\circ}$ twist angle with respect to BLG.  (b) Zoom to the calculated low energy bands (symbols) around the $K$ point, corresponding to the band structure in (a), with a fit to the model Hamiltonian (solid lines).
 [(c) and (d)] Same as [(a) and (b)] but when the top WS$_2$ layer is rotated by $60^{\circ}$ twist angle with respect to the remaining BLG/WS$_2$ structure, as schematically illustrated in (c). 
 \label{Fig:bands_TMD}}
\end{figure*}

The first case that we address is the WS$_2$ encapsulated BLG. Here, SOC is included in the calculation, since we are interested in proximity-induced SOC effects, that are present in both graphene layers. 
The calculated band structure is shown in Fig.~\ref{Fig:bands_TMD}(a). 
The dispersion features four $s_z$-polarized parabolic bands near the Fermi level, that originate from BLG and which are located within the band gaps of top and bottom TMDCs.
Because the layers interact via weak vdW forces, the dispersion of BLG stays perfectly intact. 

Zooming in to the relevant low energy bands, see Fig.~\ref{Fig:bands_TMD}(b), we find perfect agreement of the first-principles calculation results with the model Hamiltonian. The fit parameters are summarized in Table~\ref{tab:fit}, reproducing also the BLG high energy bands, splittings, and spin expectation values (see Supplemental Material~\footnotemark[1]).
We know from previous calculations \cite{Gmitra2017:PRL,Zollner2018:NJP,Zollner2020:PRL}, that a substrate below BLG induces a dipole field in BLG, similarly to an external electric field. 
In both cases, a band gap will open in the low energy dispersion. 
Here, the WS$_2$ encapsulated structure is roughly
$z$-mirror symmetric, and nearly no dipole field is present. Hence, the orbital band gap is closed. In addition, both graphene layers experience the same amount of proximity-induced valley-Zeeman type SOC, see Table~\ref{tab:fit}.
In addition, they experience almost opposite Rashba SOC, since the bottom (top) graphene layer has the WS$_2$ as substrate (capping) layer. The small difference in value arises due to slightly different interlayer distances. 
To understand this, one has to see each graphene layer individually, being encapsulated by a WS$_2$ layer and another graphene layer. 
For bare BLG, we know that the intrinsic Rashba SOC is small (5~$\mu$eV) \cite{Konschuh2012:PRB}. 
It is then the WS$_2$ layer, being responsible for strongly deforming the $p_z$-orbitals along $z$-direction, giving rise to significant Rashba SOC \cite{Gmitra2016:PRB}. 
This should not to be mixed up with the Rashba SOC originating from electric fields \cite{Gmitra2009:PRB,Konschuh2012:PRB,Konschuh2010:PRB}. 
In the Supplemental Material~\footnotemark[1], we show the calculated charge density in real space, corresponding to the low energy bands in Fig.~\ref{Fig:bands_TMD}(b). We find that the non-dimer atoms are almost exclusively responsible for the bands near the Fermi level. In addition, there are small contributions from the surrounding TMDC layers, being responsible for the sizable proximity-induced SOC. 

What happens if we now rotate the upper WS$_2$ layer by $60^{\circ}$? We will still have a fully $C_3$ symmetric heterostructure, but the layer alignment changes (remember also that we do not again relax the heterostructure after twisting). 
The most drastic consequence is in reciprocal space. 
We know that the valley-Zeeman SOC is connected to the spin-valley locking in the TMDC. 
When all layers are aligned ($0^{\circ}$ twist angles), the $K$ and $K'$ points of the individual Brillouin zones are also aligned and both graphene layers are equally proximitized. 
However, when the top WS$_2$ layer is rotated by $60^{\circ}$, the $K$ point of the BLG/WS$_2$ substrate couples now to the $K'$ point of the top WS$_2$ layer, and vice versa. 
Regarding experiment, there are now techniques to stack the layers under well controlled twist angles \cite{Nimbalkar2020:NML,Chen2016:AM}.
We want to know, what is the consequence for the low energy BLG bands?

In Fig.~\ref{Fig:bands_TMD}(c) we show the global band structure for the $60^{\circ}$ scenario. Overall, the dispersion is nearly the same as before, but the BLG bands are only $s_z$-polarized near the $K$ point. 
Away from the $K$ point, bands show an in-plane spin polarization (they appear gray in our color scale). 
In addition, if we zoom in to the fine structure, see Fig.~\ref{Fig:bands_TMD}(d), we find that the low energy bands are nearly unsplit away from the $K$ point. 
The reason is that the low energy bands are formed by non-dimer atoms $\textrm{C}_{\textrm{B1}}$ and $\textrm{C}_{\textrm{A2}}$ from the individual graphene layers \cite{Konschuh2012:PRB,McCann2013:RPP}, now experiencing Rashba and opposite valley-Zeeman SOC, see the fit parameters in Table~\ref{tab:fit}.
At the $K$ point, the bands have a clear sublattice and layer character because of the small intrinsic dipole field. The band splittings and polarizations are dominated by the intrinsic SOC parameters.
Therefore, bands are split and $s_z$-polarized.  

However, away from the $K$ point the bands are formed equally by non-dimer atoms, having opposite valley-Zeeman couplings, effectively
cancelling each other. Hence, bands are nearly unsplit and start to get in-plane polarized due to the sizable Rashba SOC, dominating the band polarizations. 
Again, we find perfect agreement of the first-principles calculation results with the model Hamiltonian, see Fig.~\ref{Fig:bands_TMD}(d), employing the fit parameters in Table~\ref{tab:fit}. The vanishing of the valley-Zeeman coupling for states away from the $K$ point is expected to decrease the spin relaxation anisotropy (spin lifetime ratio of out-of-plane to in-plane spins) to 50\%, characteristic for (residual) Rashba interactions only, deviating from the giant anisotropies predicted for large valley-Zeeman splittings \cite{Cummings2017:PRL}. The gate field should then restore the
giant anisotropies (10-100), by introducing valley-Zeeman couplings via layer polarization.

\begin{table}[htb]
\caption{\label{tab:fit} The fit parameters of the model Hamiltonian
$\mathcal{H}$ for the WS$_2$ and the CGT encapsulated BLG structures. For the WS$_2$ encapsulation, we consider the top WS$_2$ layer to be twisted by either $0^{\circ}$ or $60^{\circ}$. In the case of CGT encapsulation, we consider parallel (P) or antiparallel (AP) magnetizations of the two CGT layers. }
\begin{ruledtabular}
\begin{tabular}{l c c c c}
\multirow{2}{*}{system} & WS$_2$  & WS$_2$  & CGT  & CGT\\
 & ($0^{\circ}$) & ($60^{\circ}$) & (P) & (AP) \\
\hline 
$\gamma_0$ [eV] &  2.444 & 2.444 & 2.519 & 2.524 \\
$\gamma_1$ [eV] & 0.395 & 0.395 & 0.387 & 0.387 \\
$\gamma_3$ [eV] &  $-$0.288 & $-$0.282 & $-$0.296 & $-$0.295 \\
$\gamma_4$ [eV] &  $-$0.170 & $-$0.170 & $-$0.189 & $-$0.189 \\
$V$ [meV] &  $-$0.951 &  $-$1.103 & $-$0.495 & $-$0.423 \\
$\Delta$ [meV] &  12.331 & 13.346 & 9.303 &  9.255 \\
$\lambda_{\textrm{R1}}$ [meV] & 0.300 & 0.295 & 0 & 0 \\
$\lambda_{\textrm{R2}}$ [meV] & $-$0.277 & $-$0.222 & 0 & 0 \\
$\lambda_{\textrm{I}}^\textrm{A1}$ [meV] &  1.039 & 1.006 & 0 & 0 \\
$\lambda_{\textrm{I}}^\textrm{B1}$ [meV] &  $-$1.090 & $-$0.901 &  0 & 0 \\
$\lambda_{\textrm{I}}^\textrm{A2}$ [meV] &  1.029 &  $-$0.802 & 0 & 0 \\
$\lambda_{\textrm{I}}^\textrm{B2}$ [meV] &  $-$1.007 & 1.072 & 0 & 0 \\
$\lambda_{\textrm{ex}}^\textrm{A1}$ [meV] &  0 & 0& $-$3.447 & $-$3.254 \\
$\lambda_{\textrm{ex}}^\textrm{B1}$ [meV] &  0 & 0& $-$3.625 & $-$3.460 \\
$\lambda_{\textrm{ex}}^\textrm{A2}$ [meV] &  0 & 0 & $-$3.447 & 3.254 \\
$\lambda_{\textrm{ex}}^\textrm{B2}$ [meV] &  0 &  0& $-$3.625 & 3.460 \\
$E_D$ [meV] &  2.815  & 1.928 & 0.987 & 1.051 \\
dipole [debye] &  0.020  & 0.037 & 0.051 & 0.054 \\
\end{tabular}
\end{ruledtabular}
\end{table}

\subsection{CGT encapsulated BLG}

\begin{figure*}[htb]
	\includegraphics[width=0.99\textwidth]{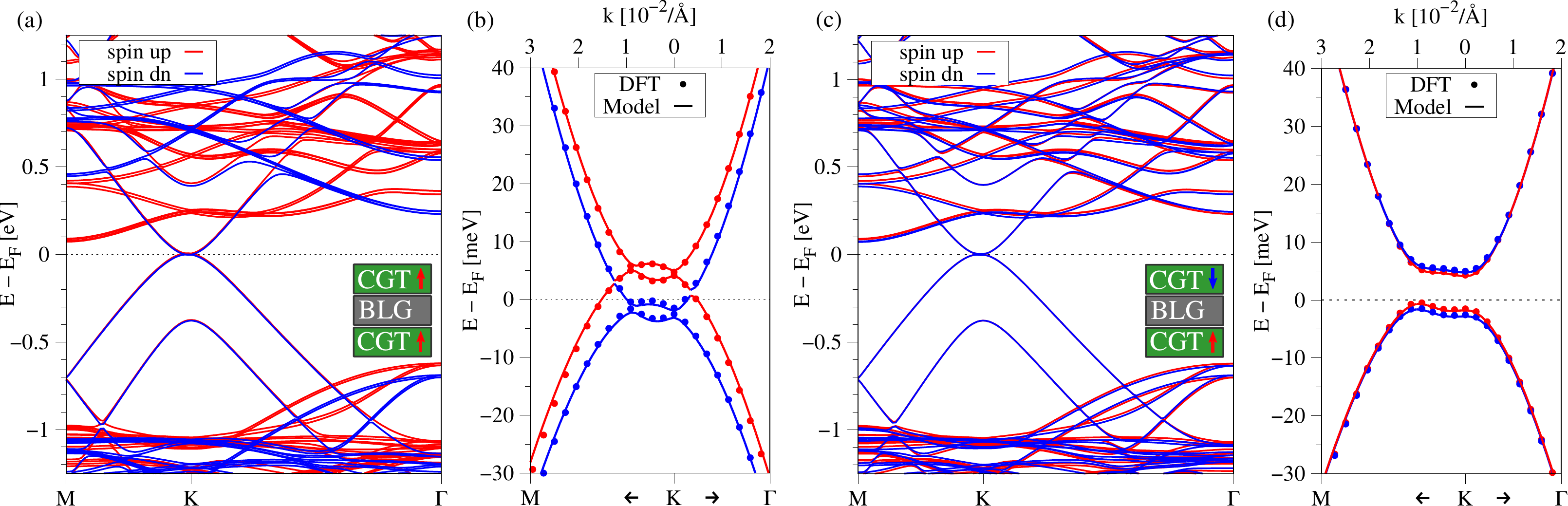}
	\caption{(a) Band structure of CGT encapsulated BLG along the $M-K-\Gamma$ path. Bands in red (blue) correspond to spin up (spin down). In the inset, we schematically illustrate the heterostructure, where both CGT layers have parallel magnetization (red arrows), pointing along the $z$ direction.
 (b) Zoom to the calculated low energy bands (symbols) around the $K$ point, corresponding to the band structure in (a), with a fit to the model Hamiltonian (solid lines). (c,d) Same as (a,b) but for antiparallel magnetization of the two CGT layers [bottom layer along $z$, top layer along $-z$, as schematically illustrated in (c)].
 \label{Fig:CGT_bands}}
\end{figure*}

Now, we consider CGT encapsulated BLG. Here, SOC is switched off in the calculation, because we are interested in bare proximity exchange parameters. In addition, we know that proximity SOC in graphene from CGT is small compared to its proximity exchange \cite{Karpiak2019:arxiv}.
Also in this case, both graphene layers are getting proximitized, and the valence and conduction band of BLG are spin split.
Depending on the magnetization directions of the two individual CGT layers, different 
low energy band structures can be realized. 
Our results show that in the ferromagnetic configuration, where the magnetizations of both CGT layers point along the $z$ direction, both graphene layers are getting equally proximitized, see Fig.~\ref{Fig:CGT_bands}(a,b). 
This is especially reflected in the model exchange parameters, $\lambda_{\textrm{ex}}$, in Table~\ref{tab:fit}, which are equal in sign and magnitude for all C sublattice atoms. 
Again, the CGT/BLG/CGT sandwich structure is 
more or less $z$-mirror symmetric and the built-in dipole field is almost zero. 
Consequently, the orbital band gap of BLG is closed, see Fig.~\ref{Fig:CGT_bands}(b), which is reflected in the small potential parameter $V$.

For the antiferromagnetic configuration, in which the magnetization of the bottom (top) CGT layer points along the $z$ ($-z$) direction, proximity exchange effects are still present, but the sign of the exchange coupling for the top graphene layer changes. 
Consequently, the spin splitting will be of similar magnitude, but of opposite sign, as also reflected in the fitted parameters in Table~\ref{tab:fit}.
Since we have switched the magnetization of one CGT layer, the global band structure is still similar to the ferromagnetic case, compare Fig.~\ref{Fig:CGT_bands}(a) and Fig.~\ref{Fig:CGT_bands}(c), but the bands corresponding to the upper CGT layer have switched their spin polarization. 
Most remarkably, the fine structure near the $K$ point reveals a sizable band gap of about 5~meV, such that no bands cross the Fermi level, see  Fig.~\ref{Fig:CGT_bands}(d).
In this case, it is not an intrinsic dipole field that is responsible for the gap opening, but rather the unique sublattice- and layer-polarized low energy band structure of BLG itself, being subject to layered antiferromagnetic proximity exchange.

Our CGT/BLG/CGT geometry is similar to the recently proposed
CrI$_3$/BLG/CrI$_3$ spin-valve heterostructure~\cite{Cardoso2018:PRL}, in which the in-plane conductance can be controlled by switching the magnetic configuration. 
Indeed, the spin-split low energy bands cross the Fermi level and the system is conductive in the ferromagnetic case, see Fig.~\ref{Fig:CGT_bands}(b), while in the antiferromagnetic case, see Fig.~\ref{Fig:CGT_bands}(d), no bands cross the Fermi level and the system is insulating. 
The advantage of our presented structure is that only the bands of the proximitized BLG are present at the Fermi level, while for the CrI$_3$/BLG/CrI$_3$ structure also bands 
originating from the CrI$_3$ layers reside near the Fermi level~\cite{Cardoso2018:PRL}.

We want to point out that this switching mechanism works for our symmetrical heterostructures, because \mbox{$V\approx0$} at zero external electric field. 
In experiment, when BLG is asymmetrically encapsulated, 
e.~g., with some additional substrate material, one needs to tune the total electric field (built-in plus external) such that $V\approx0$.

\subsection{Gate tunable low energy bands}

We now turn to the gate tunability of the low energy bands. As we know, an external electric field can be used to tune the band gap of BLG \cite{Konschuh2012:PRB,Gmitra2017:PRL,Zollner2018:NJP,Zollner2020:PRL}. 
We exploit the model Hamiltonian, along with the zero external field parameters, listed in Table~\ref{tab:fit}, to calculate the electric field behavior. 
We do not perform any additional first-principles calculations here. From previous results \cite{Zollner2020:PRL}, we know that the electric field behavior can be modelled realistically by tuning the parameter $V$ only. 
For the model calculations, we also set the Dirac point energy $E_D = 0$, since doping effects seem not to play a role, as the above first-principles results show. 

\begin{figure}[htb]
	\includegraphics[width=0.99\columnwidth]{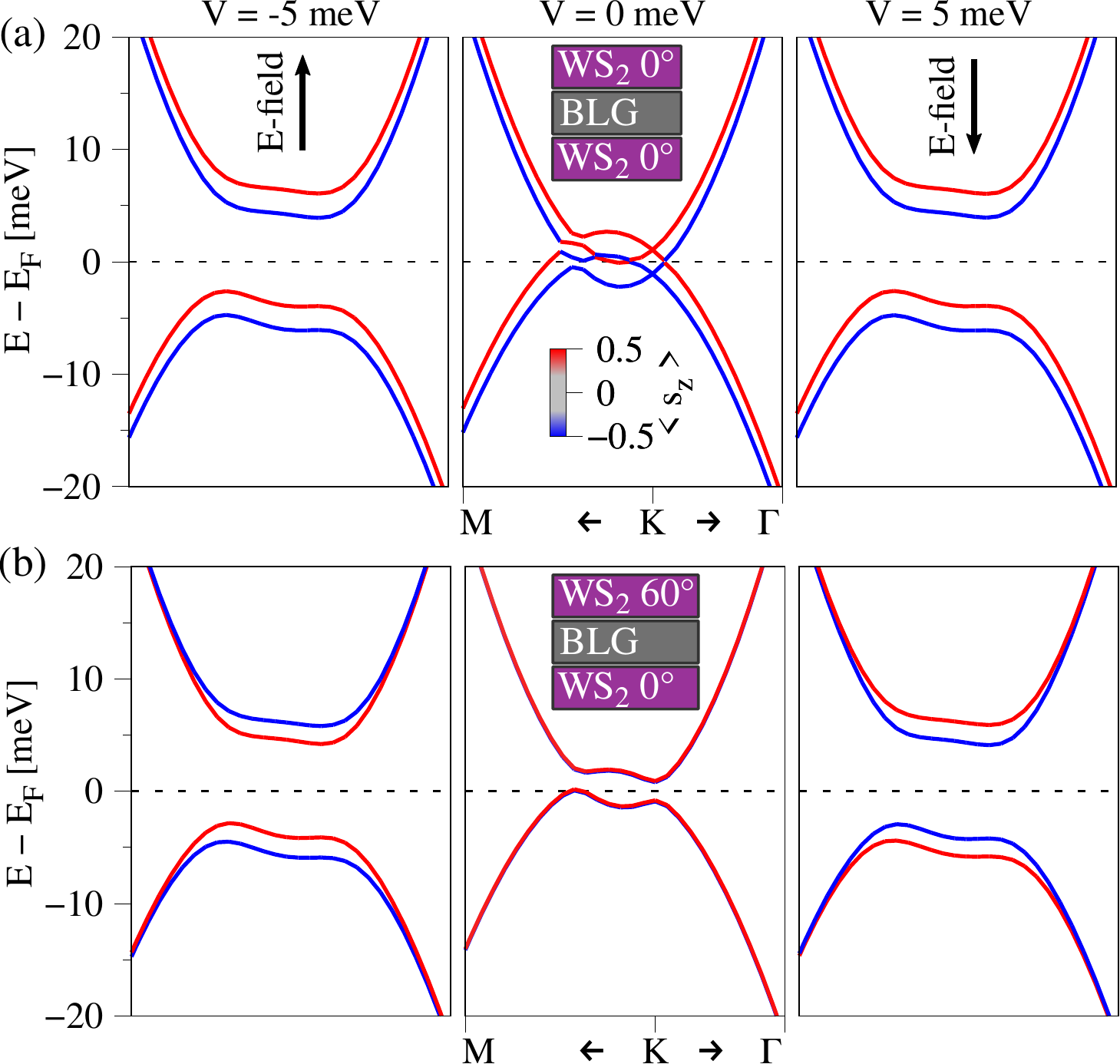}
	\caption{Calculated low energy model band structures employing the fit parameters from Table~\ref{tab:fit} for the case of WS$_2$ encapsulation, when the top WS$_2$ layer is rotated by (a) $0^{\circ}$ and (b) $60^{\circ}$. The insets schematically illustrate the situations, as in Fig.~\ref{Fig:bands_TMD}. 
	From left to right, we tune the parameter $V$ from $-5$~meV to $5$~meV, to simulate an external electric field, as indicated by the arrows. 
 \label{Fig:Model_band_series_TMD}}
\end{figure}

We restrict ourselves to three relevant cases, \mbox{$V = -5$, $0$, and $5$~meV}, corresponding to external electric fields of about $0.75$, $0$, and $-0.75$~V/nm. 
A particularly interesting case is, when $V=0$, i. e., both graphene layers are at the same potential, equally contributing to the low energy bands at the $K$ point. In our heterostructure calculations, a small asymmetry in the interlayer distances remains from the relaxation, leading to a finite intrinsic dipole and $V\neq0$, see Table~\ref{tab:fit}.

\subsubsection{WS$_2$ encapsulated BLG}

In Fig.~\ref{Fig:Model_band_series_TMD}, we summarize the model calculation results, when employing an external electric field, for WS$_2$ encapsulated BLG.
For the $0^{\circ}$ twist angle and zero electric field ($V=0$), we find that the low energy bands are equally split, $s_z$ polarized and cross the Fermi level, similar to Fig.~\ref{Fig:bands_TMD}(b).
When an electric field is applied ($V\neq 0$), a band gap opens. Independent of the field direction, the low energy bands are the same, since both graphene layers experience the same valley-Zeeman SOC. 
For the $60^{\circ}$ twist angle and zero electric field, the bands already exhibit a gap and remain nearly unsplit. The origin is that the individual valley-Zeeman SOCs from the two graphene layers are now almost opposite, nearly canceling each other. 
The presented results are in perfect agreement with recent capacitance measurements and model considerations of WSe$_2$ encapsulated BLG \cite{Island2019:Nat}. 

In general, whenever the proximity SOCs of top and bottom graphene layers cancel each other, the bands remain spin degenerate for zero electric field. 
A finite electric field opens the band gap, just as before. 
But now, depending on the field direction, \textit{the polarization of the two innermost bands can be switched, potentially important for spin-filtering purposes.} 
However, one could argue that the band splittings are small (few meV), which is a drawback for practical usage of spin filtering. 
Nevertheless, one is not limited to our scenario of WS$_2$ encapsulation and can use other 2D materials that induce a more sizable SOC in the graphene layers. 
The working principle remains the same.

\subsubsection{CGT encapsulated BLG}

\begin{figure}[htb]
	\includegraphics[width=0.99\columnwidth]{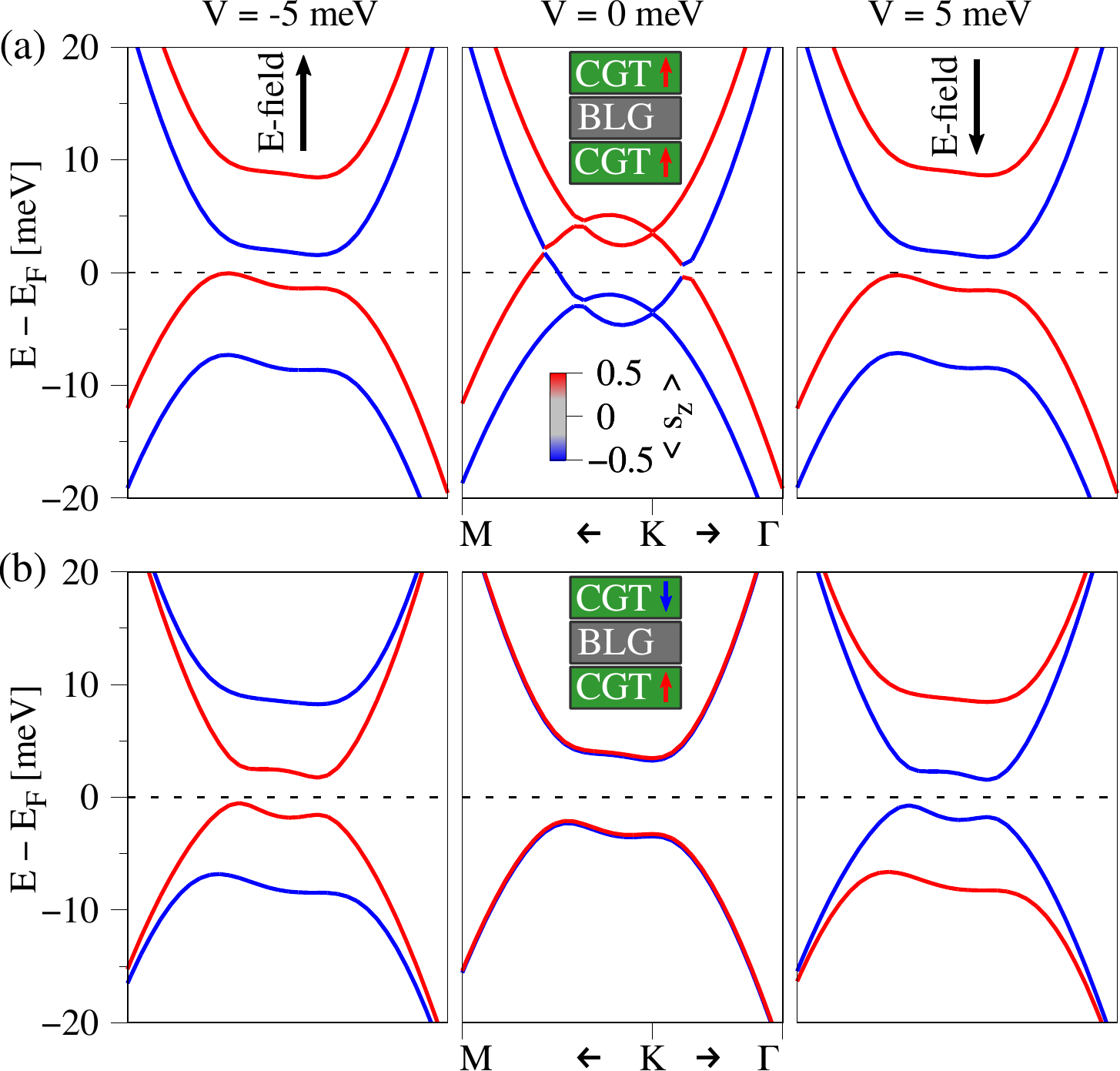}
	\caption{Calculated low energy model band structures employing the fit parameters from Table~\ref{tab:fit} for the case of CGT encapsulation, for (a) parallel (P) and (b) antiparallel (AP) magnetizations of the two CGT layers. The insets schematically illustrate the situations, as in Fig.~\ref{Fig:CGT_bands}. 
	From left to right, we tune the parameter $V$ from $-5$~meV to $5$~meV, to simulate an external electric field, as indicated by the arrows. 
 \label{Fig:Model_band_series_CGT}}
\end{figure}

In Fig. \ref{Fig:Model_band_series_CGT}, we summarize the model calculation results, when employing an external electric field, for CGT encapsulated BLG.
In the ferromagnetic case, the low energy bands can be tuned symmetrically with the field, independent of the field direction, see Fig. \ref{Fig:Model_band_series_CGT}(a). The reason is, that both graphene layers experience equal proximity effects. 
This is also, why at zero field, the bands remain split. 
In strong contrast, in the antiferromagnetic case, and at zero field, the bands are also formed equally by both graphene layers, but with opposite sign of proximity exchange. The couplings effectively cancel each other and bands remain nearly unsplit. 
Because of this layered antiferromagnetic proximity exchange, \textit{switching between positive and negative electric field allows to flip the spin-polarization of the two low energy bands}, closest to the Fermi level.   
Consequently, the relative magnetization of the two CGT layers can be used to turn on and off the in-plane conductance at zero external field. 
In the antiferromagnetic case, an external electric field can be used to control spin filtering.
Already here, we can see that the band splittings are much larger than in the WS$_2$ encapsulation case, making the CGT-encapsulated structure more favorable for practical usage of spin filtering. 
With regard to switching on and off the in-plane conductance, one could argue that the gap ($\approx 5$~meV) is rather small for practical purposes. 
However, since the size of the gap is proportional to the exchange couplings, one can simply use another ferromagnetic semiconductor that induces a more sizable proximity exchange in the graphene layers. 

In addition, we think that few meV of electronic gaps can be easily resolved now in experiments. For example recent capacitance \cite{Island2019:Nat} and magneto-transport \cite{Tiwari2021:PRL,Wang2019:NL} measurements were able to resolve such gaps in proximitized BLG and could relate this to similar band structure results that we show.

\section{Summary}

In summary, we have calculated the electronic structure of WS$_2$ or CGT encapsulated BLG from first principles. 
By employing a model Hamiltonian, we were able to reproduce the relevant low energy bands of BLG using suitable fit parameters.
The gate tunability of the bands was studied on a model level. 
Depending on the encapsulation material, either valley-Zeeman SOC or exchange coupling is induced in both graphene layers. 
Depending on the twist angle (magnetization direction) of one WS$_2$ (CGT) layer with respect to the remaining heterostructure, the corresponding proximity-induced valley-Zeeman SOC (exchange coupling) in the adjacent graphene layer can be switched in sign. 
These tunable layer-resolved proximity effects in combination with the unique sublattice character of the BLG low energy bands, allows to turn on and off the in-plane conductance at zero external electric field. 
A finite electric field can be used to further tailor the spin and charge transport properties. 

Our results show that not only proximity effects influence the low energy characteristics of BLG, but also the alignment and relative magnetization of the surrounding layers is important.  
In general, the presented findings should be also valid  when replacing the AB-stacked BLG with magic-angle twisted BLG. The splitting of the flat bands may then also be turned on and off, potentially leading to very different phase diagrams when tuning the filling factor of the Moir\'{e} Brillouin zone. This has been already demonstrated for a single WSe$_2$ substrate in proximity to twisted BLG \cite{Lin2021:arxiv,Arora2020:arxiv}.
The next step would be the encapsulation within strong spin-orbit and/or ferromagnetic semiconductors.


\acknowledgments
This work was funded by the Deutsche Forschungsgemeinschaft (DFG, German Research Foundation) SFB 1277 (Project No. 314695032), SPP 2244 (Project No. 443416183), and the European Union Horizon 2020 Research and Innovation Program under contract number 881603 (Graphene Flagship).

\footnotetext[1]{See Supplemental Material, where we show a more extended comparison between the DFT results and the model band structure, splittings, and spin expectation values. In addition, we show the calculated charge density from the low energy bands in real space.}

\bibliography{paper}

\cleardoublepage


\section*{Supplementary material}

# Supplemental Material:
# Bilayer graphene encapsulated within monolayers of WS$_2$ or Cr$_2$Ge$_2$Te$_6$: Tunable proximity spin-orbit or exchange coupling

Klaus Zollner[1, *] and Jaroslav Fabian[1]

[1]*Institute for Theoretical Physics, University of Regensburg, 93040 Regensburg, Germany*

In the Supplemental Material we again show the band structure of the WS$_2$ encapsulated BLG for the 0° twist angle case, including the model Hamiltonian fit. Here, we explicitly show that also the high energy bands, spin splittings, and spin expectation values are in good agreement with the model Hamiltonian fit. In addition, we show the DFT-calculated charge density in real space, corresponding to the low energy bands near the Fermi level.

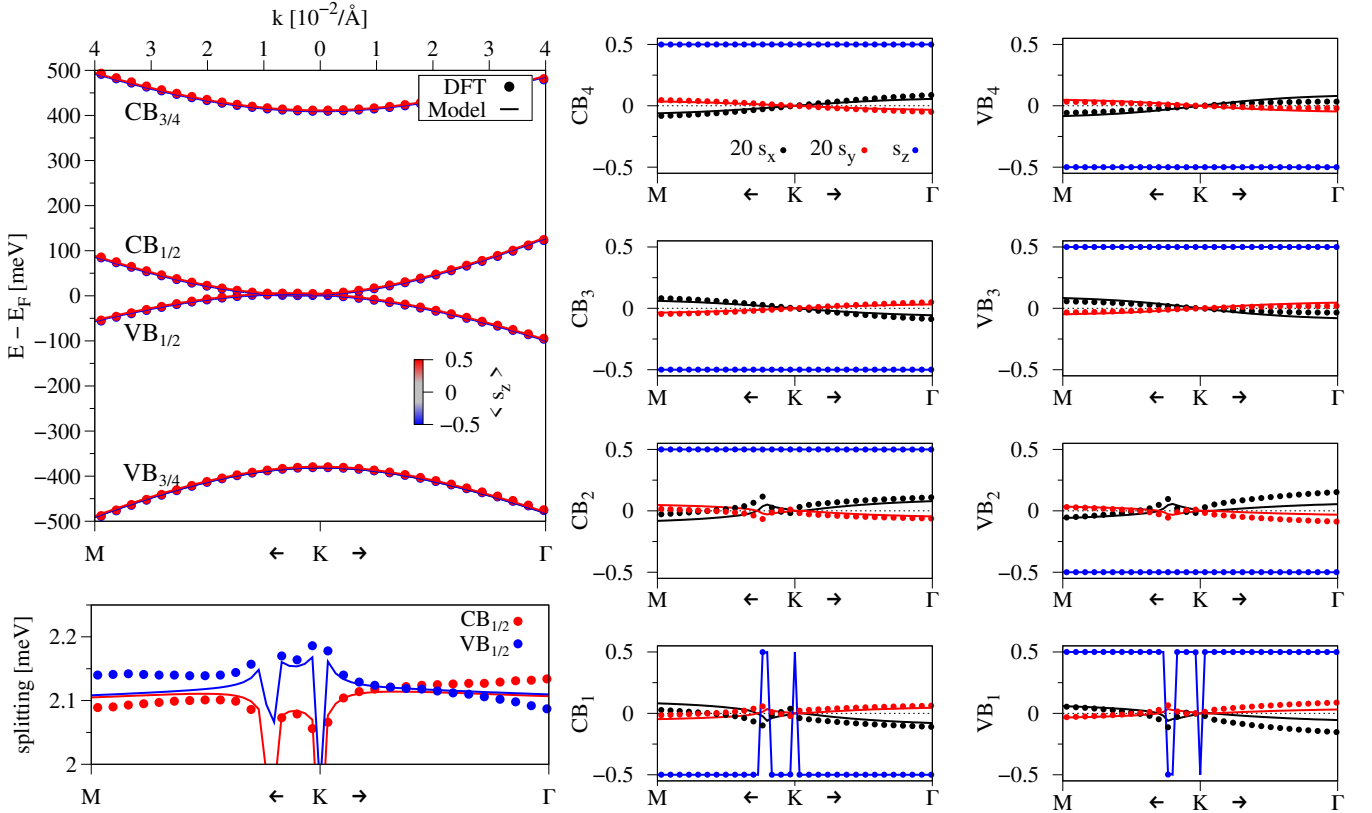

FIG. S1. DFT-calculated band structure (symbols) of WS$_2$ encapsulated BLG for the 0° twist angle case, including the fit to the model Hamiltonian (solid lines). Here, we show all 8 bands belonging to BLG, which are in good agreement with the model. The color of the bands corresponds to the $s_z$ spin expectation value. Also the band splittings and their spin expectation values ($s_x$, $s_y$, and $s_z$) agree well with the model. The in-plane spin expectation values are multiplied by a factor of 20 for better visualization.

* klaus.zollner@physik.uni-regensburg.de

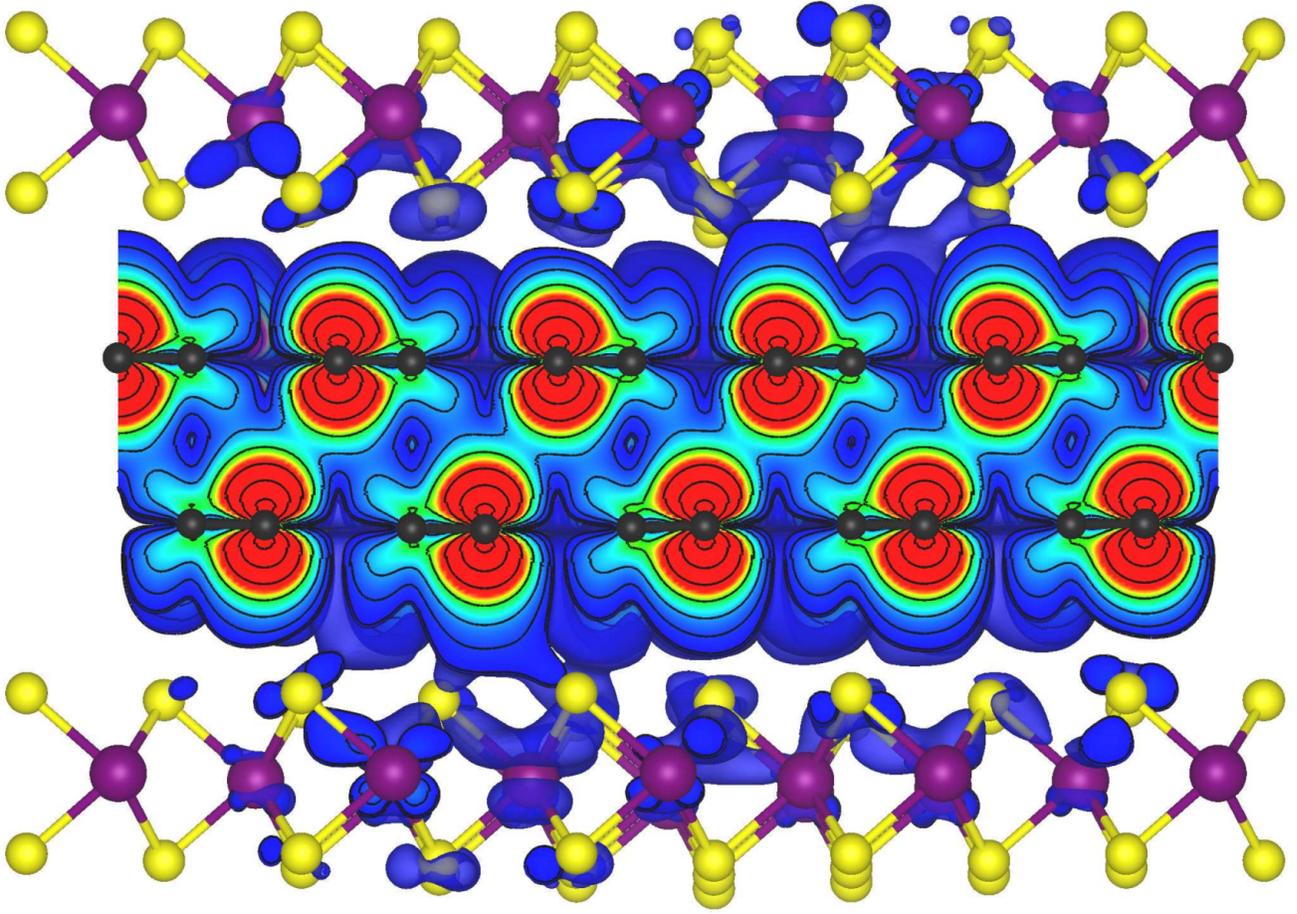

FIG. S2. DFT-calculated integrated local density of states of the WS$_2$ encapsulated BLG for the 0° twist angle case. We take into account only states in an energy window of ±25 meV around the Fermi level from the band structure in Fig. S1. The colors correspond to isovalues between $2 \times 10^{-4}$ (red) and $3 \times 10^{-6}$ (blue) e/Å$^3$, while the isolines range from $1 \times 10^{-2}$ to $1 \times 10^{-6}$ e/Å$^3$.

\clearpage

\end{document}